\documentclass[twocolumn]{aastex631}

\usepackage{multirow}
\usepackage[caption=false]{subfig}
\usepackage{graphicx}

\begin{document}

\title{Evaluating the Consistency of Cosmological Distances Using Supernova Siblings in the Near-Infrared}

\author[0000-0002-0800-7894]{Arianna M.\ Dwomoh}
\affil{Department of Physics, Duke University, Durham, NC 27708, USA}

\author[0000-0001-8596-4746]{Erik R. Peterson}
\affiliation{Department of Physics, Duke University, Durham, NC 27708, USA}

\author[0000-0002-4934-5849]{Daniel Scolnic}
\affiliation{Department of Physics, Duke University, Durham, NC 27708, USA}

\author[0000-0002-5221-7557]{Chris Ashall}
\affiliation{Department of Physics, Virginia Tech, Blacksburg, VA 24061, USA}

\author[0000-0002-7566-6080]{James M. DerKacy}
\affiliation{Department of Physics, Virginia Tech, Blacksburg, VA 24061, USA}

\author[0000-0003-3429-7845]{Aaron Do}
\affiliation{Institute for Astronomy, University of Hawai`i at M$\bar{a}$noa, Honolulu, HI 96822, USA}

\author[0000-0001-5975-290X]{Joel Johansson}
\affiliation{Oskar Klein Centre, Department of Physics, Stockholm University, AlbaNova, SE-10691 Stockholm, Sweden}

\author[0000-0002-6230-0151]{David O. Jones}
\affiliation{Gemini Observatory, NSF's NOIRLab, 670 N. A'ohoku Place, Hilo, HI 96720, USA}

\author[0000-0002-6124-1196]{Adam G. Riess}
\affiliation{Space Telescope Science Institute, Baltimore, MD 21218, USA}
\affiliation{Department of Physics and Astronomy, Johns Hopkins University, Baltimore, MD 21218, USA}

\author[0000-0003-4631-1149]{Benjamin J. Shappee}
\affiliation{Institute for Astronomy, University of Hawai`i at M$\bar{a}$noa, Honolulu, HI 96822, USA}

\begin{abstract}
The study of supernova siblings, supernovae with the same host galaxy, is an important avenue for understanding and measuring the properties of Type Ia Supernova (SN Ia) light curves (LCs). Thus far, sibling analyses have mainly focused on optical LC data. Considering that LCs in the near-infrared (NIR) are expected to be better standard candles than those in the optical, we carry out the first analysis compiling SN siblings with only NIR data. We perform an extensive literature search of all SN siblings and find six sets of siblings with published NIR photometry. We calibrate each set of siblings ensuring they are on homogeneous photometric systems, fit the LCs with the SALT3-NIR and SNooPy models, and find median absolute differences in $\mu$ values between siblings of 0.248~mag and 0.186~mag, respectively. To evaluate the significance of these differences beyond measurement noise, we run simulations that mimic these LCs and provide an estimate for uncertainty on these median absolute differences of $\sim$0.052 mag, and we find that, statistically, our analysis rules out the nonexistence of intrinsic scatter in the NIR at the 99\% level. When comparing the same sets of SN siblings, we observe a median absolute difference in $\mu$ values between siblings of 0.177~mag when using optical data alone as compared to 0.186~mag when using NIR data alone. It is unclear if these results may be due to limited statistics or poor quality NIR data; all of which will be improved with the Nancy Grace Roman Space Telescope.

\end{abstract}


\section{Introduction}\label{sec: intro}
Type Ia Supernovae (SNe Ia) are used to measure the local expansion rate of the universe \citep[e.g.,][]{Freedman2019, Riess2022}, as well as the expansion history \citep[e.g.,][]{Scolnic2022, Brout2022}. In most SN Ia cosmological analyses, SNe Ia are treated as ``standardizable candles" by using light curve (LC) properties to infer an absolute optical luminosity.
Still, after standardization, there is an unknown scatter left in luminosity magnitudes, deemed ``intrinsic scatter" ($\sigma_{\rm int}$). One way to better understand $\sigma_{\rm int}$ is by analyzing SN ``siblings," SNe found within the same host galaxy. These siblings uncover correlations between SN properties and host galaxy properties \citep{Scolnic2020, Kelsey2023}. Furthermore, siblings can be studied to help understand the effects from various local environments within a galaxy. Here, we examine SN siblings, focusing on LCs in the near-infrared (NIR).

There have been a number of SN sibling analyses using optical measurements that reach varying conclusions. 
\citet{Scolnic2020} use a Dark Energy Survey sample of photometrically-identified SNe Ia to find eight sets of siblings and compare the LC properties of those SNe to their distances. They place an upper limit that, at most, half of the $\sigma_{\rm int}$ of the SN Ia Hubble residuals can be attributed to the parent galaxy. \citet{Burns2020} find a much better distance consistency than \citet{Scolnic2020} of 3\% in a sample of 12 sets of SN Ia siblings with predominately optical LCs for observations on the same system, though they discard siblings with SWIFT telescope observations. 
\citet{Biswas2022} study a pair of siblings, found by the Zwicky Transient Facility \citep{Bellm2019} just 0.57'' apart at $z = 0.0541$, where the two SNe Ia have extremely different colors and use this to attempt to constrain the reddening law for SNe Ia. \citet{Ward2022} analyze the SN sibling trio within NGC~3147 and find that the distances to the siblings are consistent with a standard deviation smaller than the total $\sigma_{\rm int}$. \citet{Hoogendam2022} analyze a galaxy with two fast-declining SN Ia siblings and find that the distances derived from the two are consistent.

All previous SN Ia siblings analyses have reported general consistency in distance measurements in the optical, but this remains to be seen in the NIR. There is particular leverage for understanding SN scatter in the NIR with siblings, as SNe are expected to be even more uniform in luminosity in the NIR \citep{Meikle2000, Krisciunas2004, Wood-Vasey2008, Mandel2009, Folatelli2010, Phillips2012, BaroneNugent2012, Avelino2019, RAISIN,Peterson2023}. 
This uniformity is predicted because of the relationship between the optical and NIR light emitted and temperature, as explained in \citet{Kasen2006}.
As NIR LCs are less sensitive to the reddening by interstellar dust within host galaxies, NIR observations (photometric and spectroscopic alike) can also help to limit the systematic errors currently present in optical data \citep{BroutScolnic2021}. 
In this work, we perform a full literature search investigating all sets of siblings with NIR data that have been published.

In Section~\ref{sec: Data}, we describe our data collection process and our six sets of siblings. In Section~\ref{sec:data_analysis}, we detail the LC fitting process with both the SALT3-NIR and SNooPy models and compare the resulting differences in distances between siblings. In Section~\ref{sec: compare sim}, we compare results from our analysis to simulated SN Ia data and assess the implications on intrinsic scatter. We further discuss our results and present our conclusions in Section~\ref{sec: discuss and conclude}.

\begin{deluxetable*}{l @{\extracolsep{\fill}} ccclccccc}
\centering 
\tablecaption{General information about each SN sibling\label{tab:geninfo}}
\tablenum{1}
\tablehead{
\colhead{SN} &
\colhead{Host Galaxy} &
\colhead{Type} &
\colhead{NIR Photometry Source} &
\colhead{$z^a$} &
\colhead{Bands} &
\colhead{RA} &
\colhead{DEC} &
\colhead{Opt.~Peak MJD} &
\colhead{Epochs$^b$} 
}
\startdata
$\rm 1980N$ & \multirow{4}{*}{$\rm{NGC}~1316$} & $\rm Ia$ & \cite{Elias1981} & \multirow{4}{*}{$0.006010(10)$} & $\rm \textit{JH}$ & 03:23:00.30 & $-$37:12:50.00 & $44586.8(2)$ & 27 \\
$\rm 1981D$ & & $\rm Ia$ & \cite{Elias1981} & & $\rm \textit{JH}$ & 03:22:38.40 & $-$37:13:58.00 & $44680.5(2)$ & 14 \\
$\rm 2006dd$ & & $\rm Ia$ & \cite{Stritzinger2010} & & $\rm \textit{JH}$ & 03:22:41.62 & $-$37:12:13.00 & $53919.5(1)$ & 44 \\
$\rm 2006mr$ & & 91bg-like & \cite{Contreras2010} & & $\rm \textit{JH}$ & 03:22:42.84 & $-$37:12:28.51 & $54050.4(1)$ & 59 \\
\hline
$\rm 2002bo$ & \multirow{2}{*}{$\rm{NGC}~3190$} & $\rm Ia$ & \cite{Benetti2004} & \multirow{2}{*}{$0.004370(13)$} & $\rm \textit{JH}$ & 10:18:06.51 & $+$21:49:41.70 & $52356.0(2)$ & 10 \\
$\rm 2002cv$ & & $\rm Ia$ & \cite{Elias-Rosa2008} & & $\rm \textit{JH}$ & 10:18:03.68 & $+$21:50:06.00 & $52419.6(7)^c$ & 27 \\
\hline
$\rm 2002dj$ & \multirow{2}{*}{$\rm{NGC}~5018$} & $\rm Ia$ & \cite{Pignata2008} & \multirow{2}{*}{$0.009393(30)$} & $\rm \textit{JH}$ & 13:13:00.34 & $-$19:31:08.69 & $52451.1(1)$ & 42 \\
$\rm 2021fxy$ & & $\rm Ia$ & DEHVILS & & $\rm \textit{YJH}$ & 13:13:01.57 & $-$19:30:45.18 & $59306.1(2)$ & 18 \\
\hline
$\rm 2007on$ & \multirow{2}{*}{$\rm{NGC}~1404$} & $\rm Ia$ & CSP & \multirow{2}{*}{$0.006494(13)$} & $\rm \textit{YJH}$ & 03:38:50.90 & $-$35:34:30.00 & $54420.5(1)$ & 155 \\
$\rm 2011iv$ & & $\rm Ia$ & CSP & & $\rm \textit{YJH}$ & 03:38:51.35 & $-$35:35:31.99 & $55905.3(1)$ & 58 \\
\hline
$\rm 2011at$ & \multirow{2}{*}{MCG -2-24-27} & $\rm Ia$ & CfA & \multirow{2}{*}{$0.006758(2)$} & $\rm \textit{JH}$ & 09:28:57.56 & $-$14:48:20.59 & $55635.0(50)^d$ & 27 \\
$\rm 2020jgl$ & & $\rm Ia$ & DEHVILS & & $\rm \textit{YJH}$ & 09:28:58.43 & $-$14:48:19.88 & $58993.0(3)$ & 15 \\
\hline
$\rm 2020sjo$ & \multirow{2}{*}{$\rm{NGC}~1575$} & $\rm Ia$ & DEHVILS & \multirow{2}{*}{$0.031265(21)$} & $\rm \textit{YJH}$ & 04:26:21.95 & $-$10:05:55.72 & $59107.6(2)$ & 20 \\
$\rm 2020zhh$ & & $\rm Ia$ & DEHVILS & & $\rm \textit{YJH}$ & 04:26:19.84 & $-$10:05:56.86 & $59178.4(2)$ & 3 \\
\enddata
\tablecomments{\\$^a$ Heliocentric redshift of the host galaxy from NED.\\$^b$ Total number of NIR observations in the bandpasses used in this analysis. Each filter is counted separately. \\$^c$ We fit for the optical peak MJD for this SN finding 52418.1(2) with SALT3-NIR and 52419.1(1) with SNooPy.\\$^d$ We fit for the optical peak MJD for this SN finding 55631.3(6) with SALT3-NIR and 55628.2(4) with SNooPy.}
\end{deluxetable*}

\section{Data Sample}\label{sec: Data}
\subsection{Data Collection}
Outside of the Carnegie Supernova Project \citep[CSP; ][]{Hamuy2006}, CfA \citep{Wood-Vasey2008, Friedman2015}, and the Dark Energy, H$_0$, and peculiar Velocities using Infrared Light from Supernovae \citep[DEHVILS; ][]{Peterson2023} survey, there are few large samples of publicly available high-quality NIR data. We used existing (optical) sibling analyses in order to construct a list of SNe for which to search for NIR data. \citet{Kelsey2023} has a large sample of siblings compiled, totaling $>$150 sets of SN siblings. 
We also searched on the Transient Name Server\footnote{\url{https://www.wis-tns.org/}.} for any SNe within 2' of the DEHVILS SNe detailed in \citet{Peterson2023} looking for potential novel pairs of SN siblings not already listed in previous works.

After compiling this list of all SN siblings in the literature, we used it to look for as many siblings with published NIR photometric data as possible. In total, we found six sets that had any NIR photometric coverage for both siblings.

\subsection{Siblings}
The six sets of siblings we analyze in this paper are SN 1980N/1981D/2006dd/2006mr, SN 2002bo/2002cv, SN 2002dj/2021fxy, SN 2007on/2011iv, SN 2011at/2020jgl, and SN 2020sjo/2020zhh. Information on these SN siblings is further detailed in Table~\ref{tab:geninfo}. For each SN, we list its host galaxy, type, information source, heliocentric redshift from references linked in the NASA/IPAC Extragalactic Database (NED),\footnote{\url{https://ned.ipac.caltech.edu/}.} NIR filter bandpasses used in this analysis, coordinates, Modified Julian Date (MJD) of the time of peak in the optical, and the total data points for the NIR LC. 
Optical peak MJD values for DEHVILS SNe are given in \citet{Peterson2023} and calculated from fits to ATLAS \citep{ATLAS} LCs.
We use the peak MJD values from Pantheon+ \citep{Scolnic2022,Brout2022} whenever available; otherwise, we use the reported values from the cited papers in Table~\ref{tab:geninfo}.
$J$-band images of the host galaxies are displayed in Fig.~\ref{fig: cutouts} with the location of each of the siblings indicated as well. 
Images with a DEHVILS SN in them are template images from the DEHVILS survey \citep{Peterson2023}, and those without a DEHVILS SN are from the Two Micron All Star Survey \citep[2MASS; ][]{Skrutskie2006}.
% All stellar photometry from each LC source in this sample is analyzed in Appendix~\ref{appendix_photometric_analysis} where we detail our calibration efforts, comparing stellar photometry with that from 2MASS.
% We compare all stellar photometry from each LC source to that from 2MASS in Appendix~\ref{appendix_photometric_analysis} where we .}
In the following sections we describe the photometry used for each SN sibling in this analysis.
All of the photometric data detailed here have been compiled and are available online at \url{https://github.com/ariannadwomoh/SNIRS}.

\newcommand{\figonewidth}{0.4}
\begin{figure*}[!htb]
\centering
\subfloat{
    \includegraphics[width=\figonewidth\textwidth]{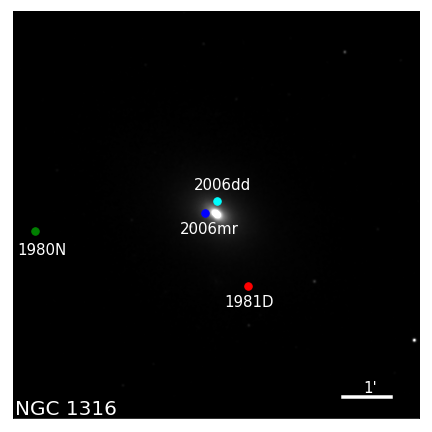}
    \label{fig:eee}\qquad
}
\subfloat{
    \includegraphics[width=\figonewidth\textwidth]{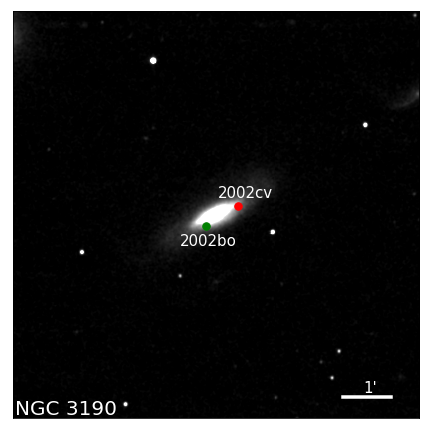}
    \label{fig:fff}\qquad
}
\\
\subfloat{
    \includegraphics[width=\figonewidth\textwidth]{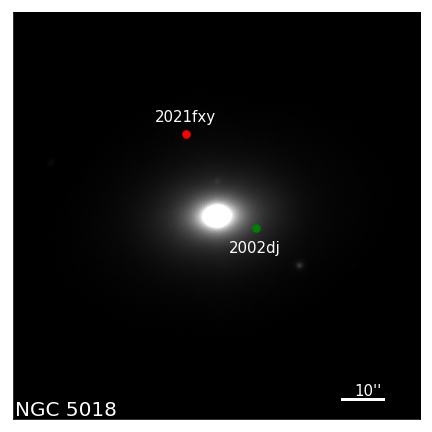}
    \label{fig:aaaa}\qquad
}
\subfloat{
    \includegraphics[width=\figonewidth\textwidth]{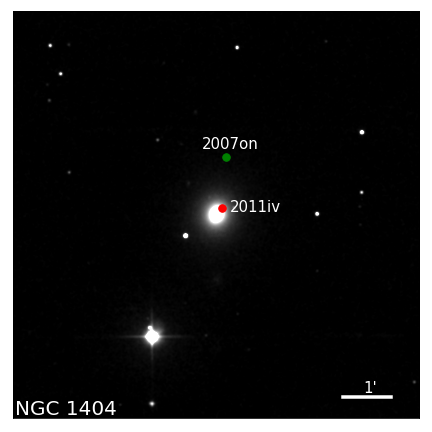}
    \label{fig:dddd}\qquad
}
\\
\subfloat{
    \includegraphics[width=\figonewidth\textwidth]{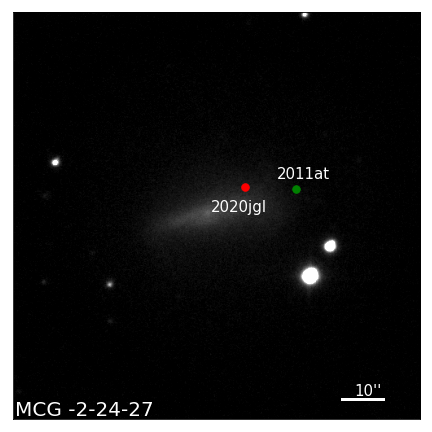}
    \label{fig:bbbb}\qquad
}
\subfloat{
    \includegraphics[width=\figonewidth\textwidth]{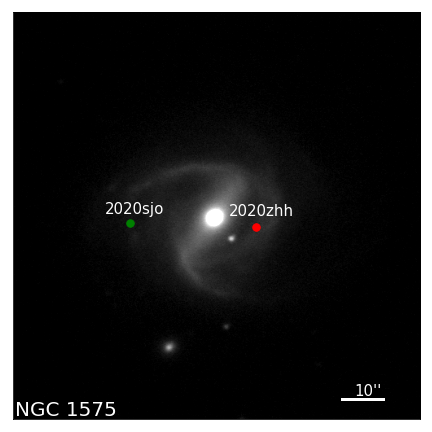}
    \label{fig:cccc}\qquad
}
\caption{Images of the six SN sibling host galaxies. The colored markers indicate the locations of the SNe. All images are in $J$-band, and North is up and East is to the left for each image.} 
\label{fig: cutouts}
\end{figure*}

\subsubsection{\rm{SN~1980N, SN~1981D, SN~2006dd, \& SN~2006mr}}
This sibling set has four SNe within the host galaxy NGC 1316, as seen in Fig.~\ref{fig: cutouts}. All SNe are Type Ia, with SN~2006mr \citep{Monard2006mr} being 91bg-like. The NIR photometry for SN~1980N \citep{Maza1980N} and SN~1981D \citep{Cragg1981D, Evans1981D} is given in \citet{Elias1981}, and the observations in the \textit{\textit{JHK}}-bands were obtained using the Cerro Tololo Inter-American Observatory (CTIO) 4-meter and 1.5-meter telescopes. For SN~2006dd \citep{Monard2006dd}, \citet{Stritzinger2010} obtained their photometry at CTIO using the SMARTS 1.3-meter telescope equipped with A Novel Double-Imaging CAMera (ANDICAM) and a small portion of the data with the Swope telescope and RetroCam imager as a part of CSP in the \textit{JH$K_{s}$}-bands. \citet{Contreras2010} used the Irenee du Pont (du Pont) 2.5-meter telescope with the Wide Field IR Camera (WIRC) to obtain photometry for SN~2006mr in the \textit{YJH}-bands.

\textit{JH$K_{s}$} magnitudes for SN~2006dd were calibrated directly to the \citet{Persson1998} NIR photometric system. Similar to SN~2006dd, the local NIR sequence for SN~2006mr was calibrated to the \citet{Persson1998} system. \citet{Elias1981} state that the observations for both SN~1980N and SN~1981D are on the natural \textit{JHKL} system of their respective observatories.

\subsubsection{\rm{SN~2002bo \& SN~2002cv}}
This sibling pair, SN~2002bo \citep{Cacella2002bo} and SN~2002cv \citep{Larionov2002cv}, is located within NGC 3190. The NIR data for SN~2002bo, as detailed in \citet{Benetti2004}, were obtained using the 1.55-meter Carlos Sanchez Telescope with the CAIN infrared camera as well as the 2.5-meter Nordic Optical Telescope (NOT) with the NOTCam infrared camera in the \textit{JH$K_{s}$}-bands. SN~2002cv NIR data in the \textit{JH$K_{s}$}-bands were found using the 1.08-meter AZT-24 Telescope along with SWIRCAM and the 3.6-meter European Southern Observatory New Technology Telescope (ESO-NTT) equipped with the Son of ISAAC (SofI) camera, as noted in \citet{Elias-Rosa2008}.

For SN~2002bo, all observations are calibrated using NOTCam images and the photometric standards stated in \citet{Hunt1998} and \citet{Persson1998}. 
The calibration of SN~2002cv data was derived using two local stars.

Of note, we use the more tightly constrained value for the optical peak MJD value for SN~2002bo from \citet{Krisciunas2002bo} with an uncertainty of 0.2 days rather than that from \citet{Benetti2004} in Table~\ref{tab:geninfo}.
Additionally, since SN~2002cv was not observable in the $B$-band, \citet{Elias-Rosa2008} report an $I$-band maximum of $\textrm{MJD}=52414.6 \pm 0.2$.
We make an initial estimate for $B$-band maximum as approximately 5 days later than $I$-band maximum and incorporate extra uncertainty on the value.
Because of this, we fit for a time of maximum when fitting the LC from SN~2002cv.

\subsubsection{\rm{SN~2002dj \& SN~2021fxy}}
We obtain NIR photometric data in the \textit{JH}- and \textit{YJH}-bands for SN~2002dj \citep{Hutchings2002dj} and SN~2021fxy \citep{Itagaki2021fxy}, found in NGC 5018 \citep{DerKacy2023}, from \citet{Pignata2008} and DEHVILS, respectively. For SN~2002dj, the NIR data were obtained using the 1.0-meter telescope at CTIO equipped with ANDICAM and ESO-NTT equipped with SofI. For SN~2021fxy, the Wide Field CAMera (WFCAM) mounted onto the 3.8-meter United Kingdom Infrared Telescope (UKIRT) was used.

NIR data reductions for SN~2002dj were carried out using standard Image Reduction and Analysis Facility (IRAF) routines. An illumination correction was applied to all SofI images and four stars close to SN~2002dj were calibrated in the \textit{JH$K_{s}$}-bands using the \citet{Persson1998} system. For the calibration of SN~2021fxy, and for all other DEHVILS SNe, \citet{Peterson2023} do a transformation of 2MASS magnitudes to WFCAM magnitudes which are further calibrated using Hubble Space Telescope CALSPEC stars.

\subsubsection{\rm{SN~2007on \& SN~2011iv}}
The next pair of siblings are located in NGC 1404, and they are SN~2007on \citep{Pollas2007on} and SN~2011iv \citep{Drescher2011at}. All NIR data in the \textit{YJH}-bands were detailed in \citet{Gall2018} as a part of CSP. NIR imaging of SN~2007on was taken using the Swope telescope using RetroCam and the du Pont telescope with WIRC. As for SN~2011iv, data were collected using the du Pont telescope with RetroCam. 

\citet{Gall2018} state that the \textit{JH}-band values were calibrated to the \citet{Persson1998} system while the $Y$-band was calibrated to standard $Y$-band magnitudes computed using stellar atmosphere models \citep{Castelli2003} along with the \citet{Persson1998} system, as done in \citet{Hamuy2006}.

\subsubsection{\rm{SN~2011at \& SN~2020jgl}}
SN~2011at \citep{Cox2011at} and SN~2020jgl \citep{Tonry2020jgl} were found in MCG -2-24-27. \citet{Friedman2015} used the 1.3-meter Peters Automated InfraRed Imaging TELescope (PAIRITEL) at the Fred Lawrence Whipple Observatory to obtain \textit{JH$K_{s}$}-band data for SN~2011at with CfA. \citet{Peterson2023} state that the NIR data for SN~2020jgl were obtained in the \textit{YJH}-bands using WFCAM with DEHVILS.

PAIRITEL's photometric observations for SN~2011at are calibrated with the 2MASS field star observations. SN~2020jgl LC data are from DEHVILS, and they are calibrated as such. Of note, there is no optical data and the NIR data obtained for SN~2011at covers only the secondary maximum of the LC (all data are $\gtrsim$ 10 days past optical maximum). Because of this, we fit for a time of maximum when fitting the LC from SN~2011at.

\subsubsection{\rm{SN~2020sjo \& SN~2020zhh}}
All NIR data for the siblings SN~2020sjo \citep{Perley2020sjo} and SN~2020zhh \citep{Chambers2020zhh}, which are located in NGC 1575, were noted in \citet{Peterson2023} and obtained in the \textit{YJH}-bands using WFCAM. 

Since both siblings are from DEHVILS, they are calibrated in the same way with 2MASS and a refinement from CALSPEC stars.

\newcommand{\mywidth}{0.49}
\begin{figure*}[!htb]
\subfloat{
    \includegraphics[width=\mywidth\textwidth]{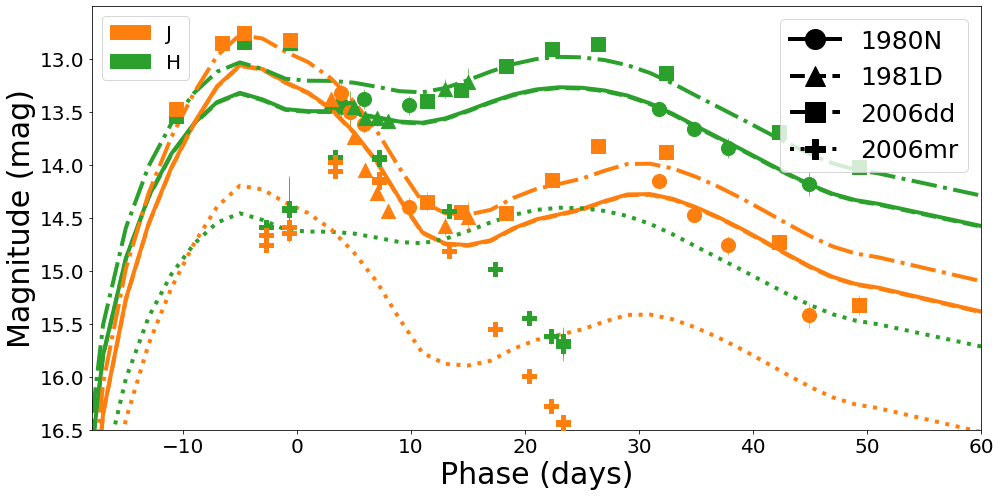}
    \label{fig:ggg}\qquad
}
\hspace{-0.8cm}
\subfloat{
    \includegraphics[width=\mywidth\textwidth]{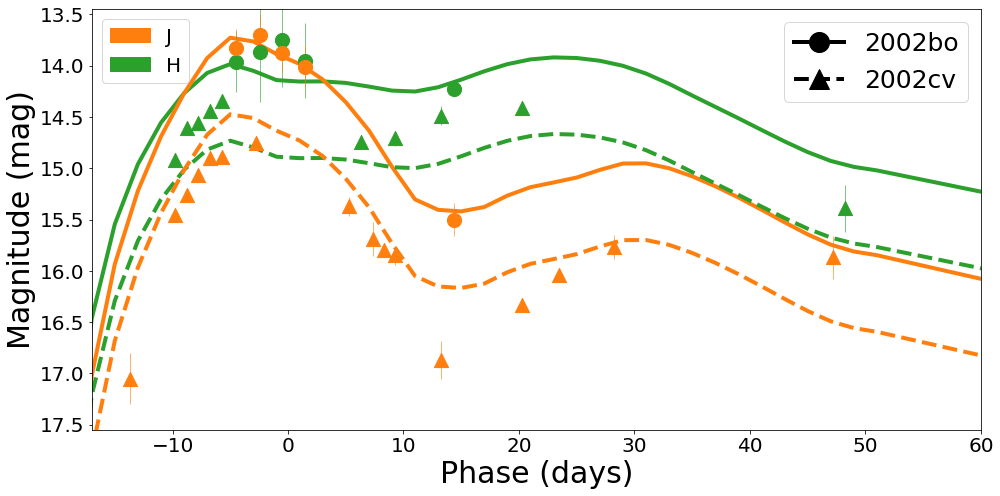}
    \label{fig:hhh}\qquad
}
\\
\subfloat{
    \includegraphics[width=\mywidth\textwidth]{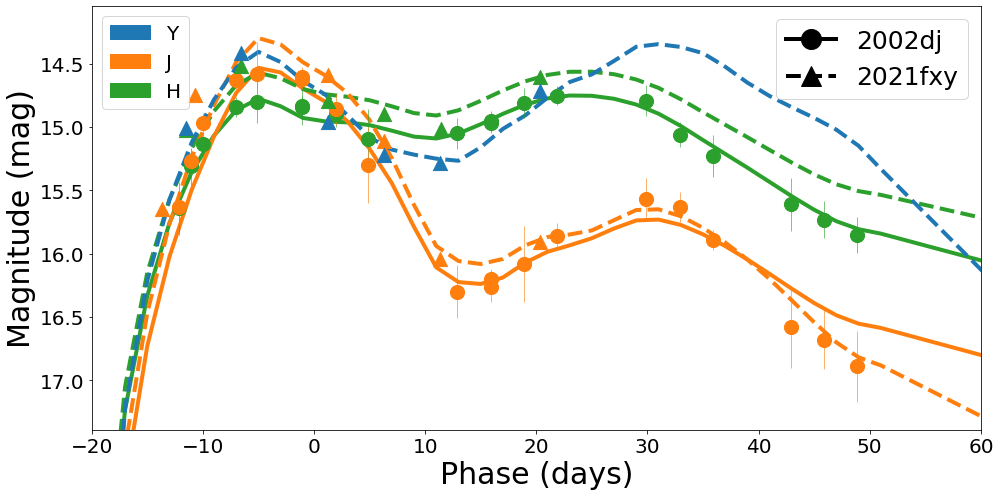}
    \label{fig:iii}\qquad
}
\hspace{-0.75cm}
\subfloat{
    \includegraphics[width=\mywidth\textwidth]{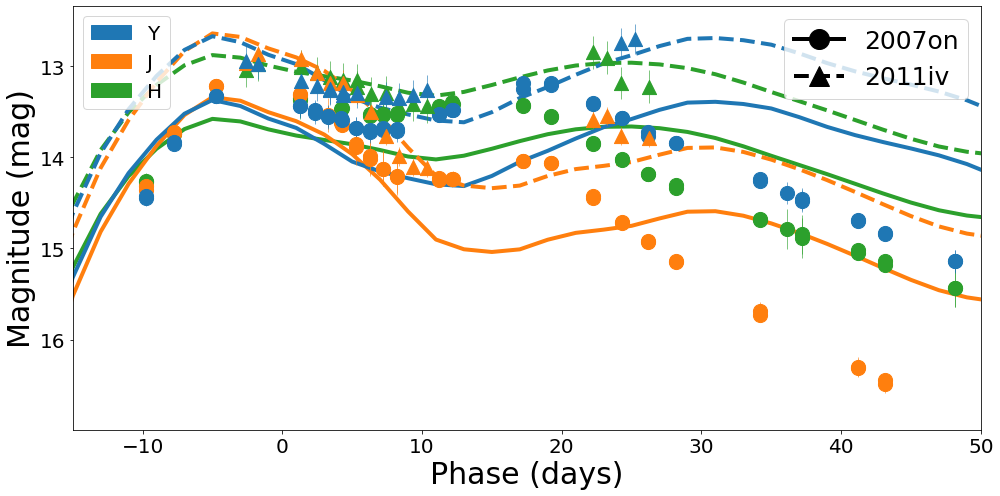}
    \label{fig:jjj}\qquad
}
\\
\subfloat{
    \includegraphics[width=\mywidth\textwidth]{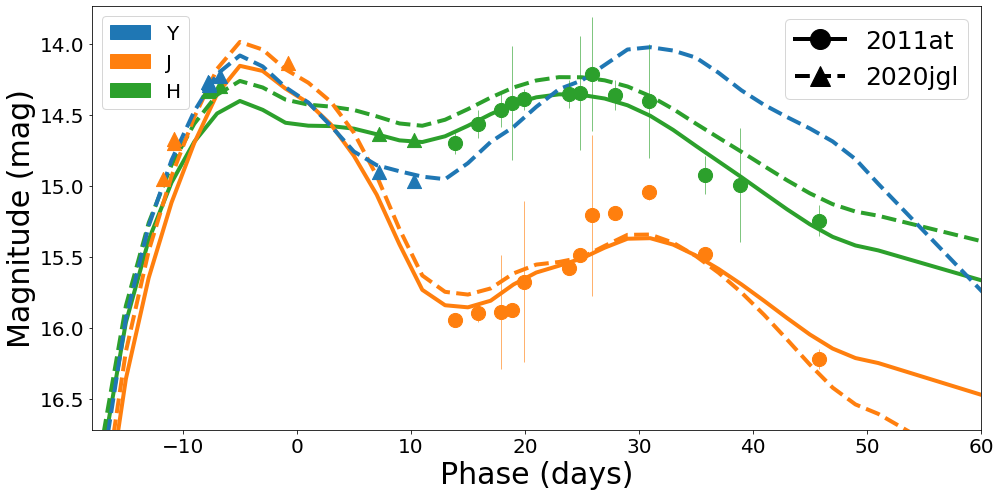}
    \label{fig:kkk}\qquad
}
\hspace{-0.8cm}
\subfloat{
    \includegraphics[width=\mywidth\textwidth]{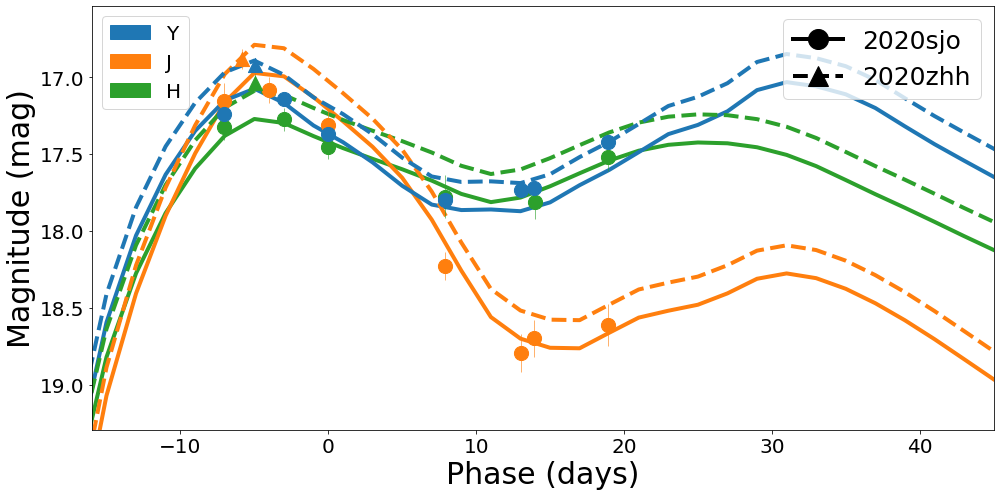}
    \label{fig:lll}\qquad
}
\caption{LC fits for each sibling set in our sample using SALT3-NIR. Phase is given with respect to optical peak. Each subplot features a specific marker and line style for every SN. The band colors are the same for all SNe and all panels.} 
\label{fig:salt3 lightcurves}
\end{figure*}

\section{Data Analysis}\label{sec:data_analysis}

\begin{figure*}[!htb]
\subfloat{\includegraphics[width=\mywidth\textwidth]{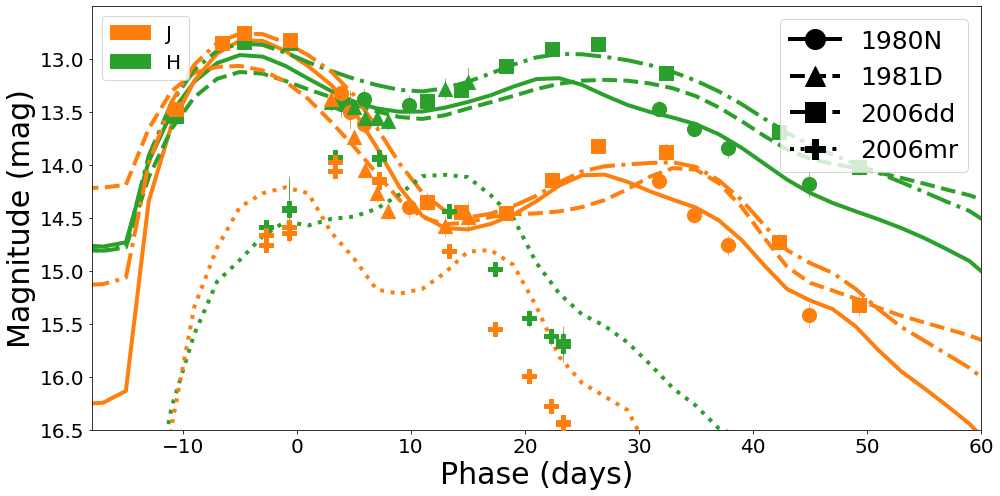}
    \label{fig:mmm}\qquad
}
\hspace{-0.8cm}
\subfloat{\includegraphics[width=\mywidth\textwidth]{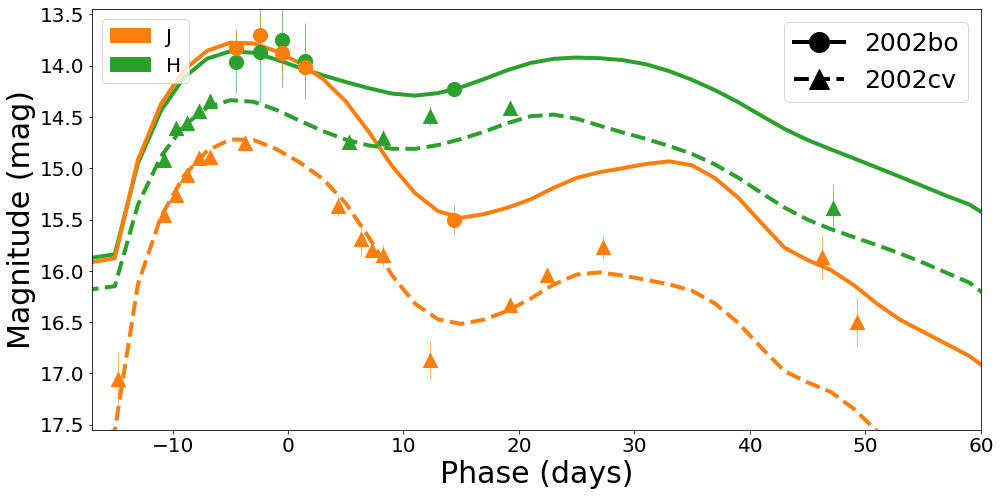}
    \label{fig:nnn}\qquad
}
\\
\subfloat{\includegraphics[width=\mywidth\textwidth]{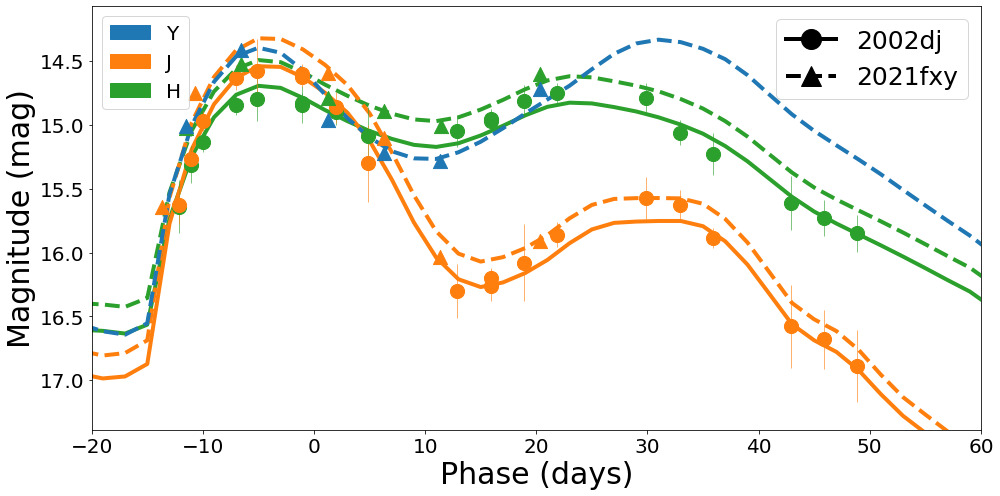}
    \label{fig:ooo}\qquad
}
\hspace{-0.7cm}
\subfloat{\includegraphics[width=\mywidth\textwidth]{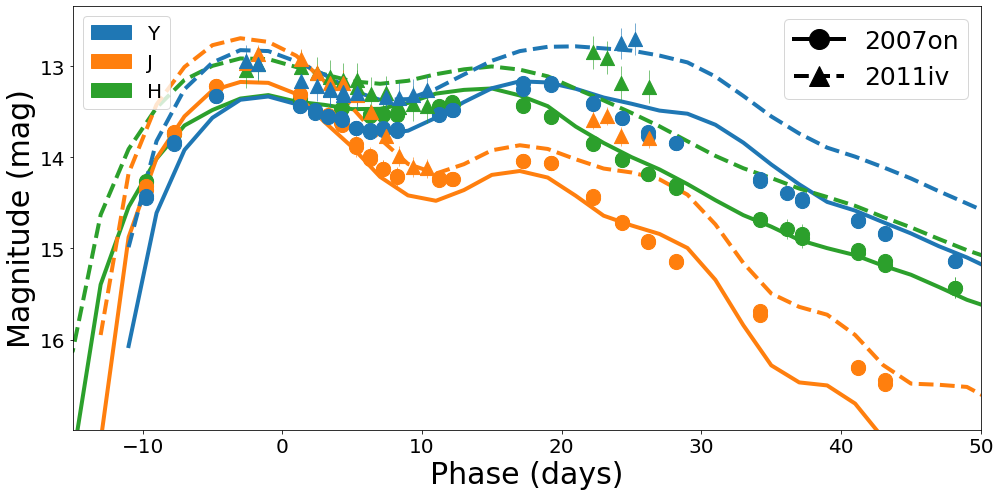}
    \label{fig:ppp}\qquad
}
\\
\subfloat{\includegraphics[width=\mywidth\textwidth]{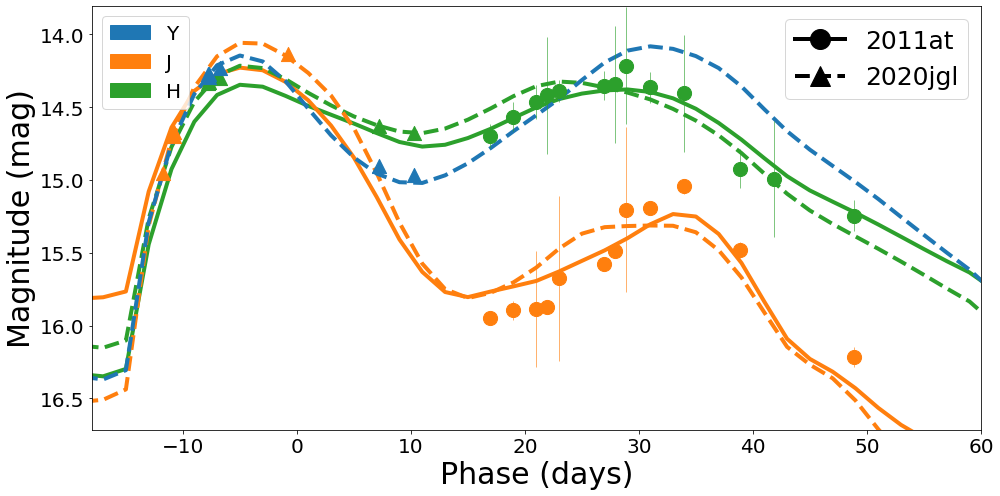}
    \label{fig:qqq}\qquad
}
\hspace{-0.8cm}
\subfloat{\includegraphics[width=\mywidth\textwidth]{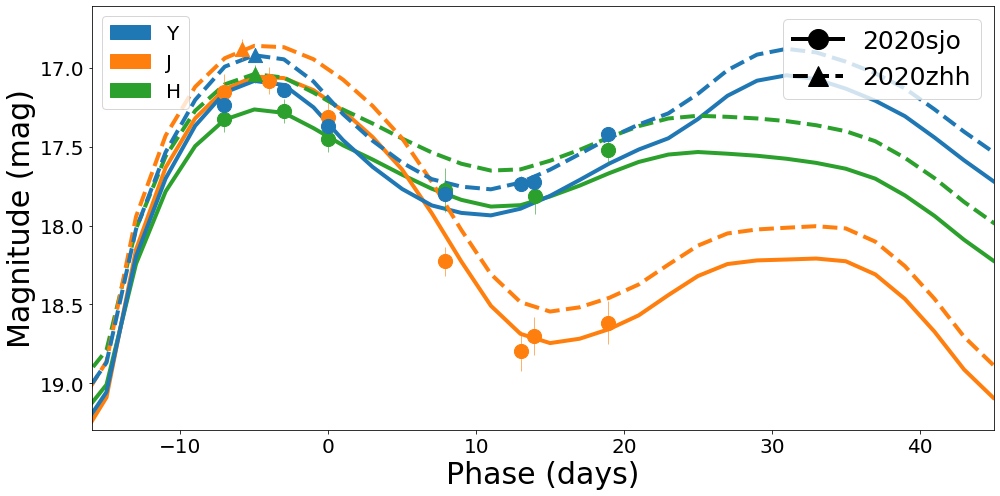}
    \label{fig:rrr}\qquad
}
\caption{Similar to Fig.~\ref{fig:salt3 lightcurves} but with LC fits using SNooPy.} 
\label{fig:snoopy lightcurves}
\end{figure*}

\subsection{Light Curve Fits}\label{sec: lc fits}
We use the SuperNova ANAlysis (\texttt{SNANA}) software package to both fit and simulate SN Ia LCs \citep{Kessler2009}. In order to measure the distance to each SN, we fit LCs using both the Spectral Adaptive Light curve Template \citep[SALT3-NIR;][]{Kenworthy2021,Pierel2022} and SuperNova object oriented Python \citep[SNooPy;][]{Burns2011, Burns2014} EBV\_model2 LC models. 
With the exception of SN LCs from CSP and DEHVILS, which have well understood photometric systems,
almost every SN LC is obtained with a different telescope and instrument.
Due to the challenge of deducing the filter functions used and the likelihood that all data were transformed to a similar 2MASS system, we treat each photometric system as if it were that from 2MASS. 
As discussed in Appendix~\ref{appendix_photometric_analysis}, this is because we find that the calibration of the stellar observations around each SN is relatively consistent with the 2MASS system. 
While the range of median magnitude residuals is from $-$0.045 to 0.130~mag, the median difference between 2MASS photometry and all other stellar photometry compiled is 0.008 mag.

SALT3-NIR fits from all six sets of siblings are presented in Fig.~\ref{fig:salt3 lightcurves}, and SNooPy fits are presented in Fig.~\ref{fig:snoopy lightcurves}. Fits to the data are capped at 50 days past optical peak, and the peak MJD value is fixed to the optical peak value for each SN given in Table~\ref{tab:geninfo} (with the exceptions of SN~2002cv and SN~2011at where we fit for the time of maximum because their peaks are both ill-constrained).
All fits from SALT3-NIR have both the stretch parameter, $x_1$, and color parameter, $c$, fixed to zero (discussed further in Section~\ref{subsec:mu_comparisons}) such that each SN NIR LC is treated as a standard candle, only the scaling parameter $x_0$ is fit, and each $\mu$ value is given by,

\begin{equation}
    \mu = m_B - \mathcal{M},
\end{equation}

\noindent where $m_B = -2.5\log(x_0) + const.$ and $\mathcal{M}$ is the globally fit absolute peak magnitude derived for the set of SNe and is fixed as a single value for all siblings.

SNooPy fits, on the other hand, incorporate stretch and color corrections.
SNooPy's stretch parameter, $s_{BV}$, and $A_V$ are fit for in all cases except in extreme cases such as fast-declining ($s_{BV}<0.7$) or highly-extincted ($A_V>5.0$~mag) events because \texttt{SNANA} SNooPy minimization fails at these levels. 
For example, for SN~2002cv, we fix $A_V$ to the derived value of 8.4~mag reported in \citet{Elias-Rosa2008}, and for SN~2011iv and SN~2007on, we fix $s_{BV}$ to the reported values of 0.64 and 0.57, respectively, from \citet{Gall2018}.

\subsection{$\mu$ Comparisons}\label{subsec:mu_comparisons}

\begin{figure}[!htb] 
\centering
\subfloat{\includegraphics[width=\mywidth\textwidth]{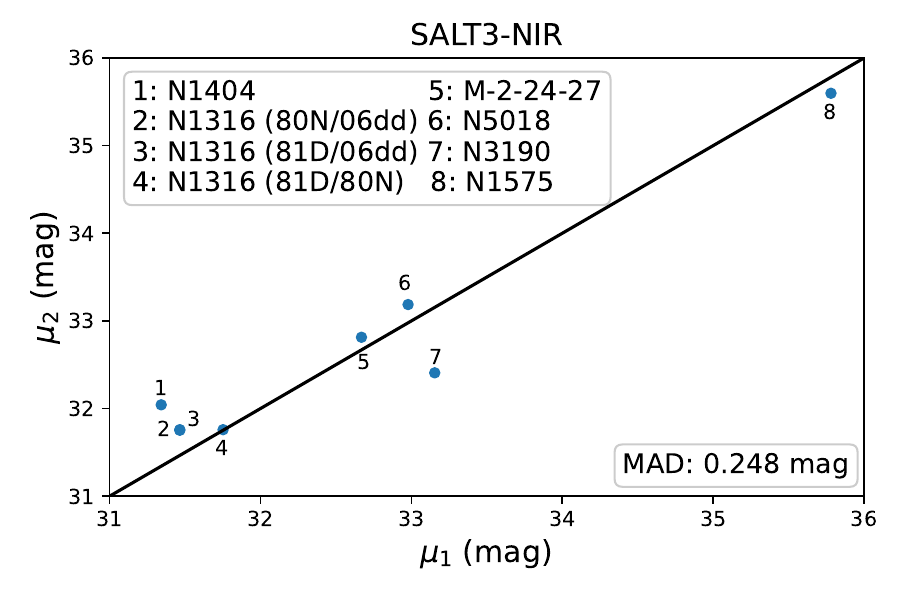}
        \label{fig:mu vs mu salt3}\qquad
}
\\
\subfloat{\includegraphics[width=\mywidth\textwidth]{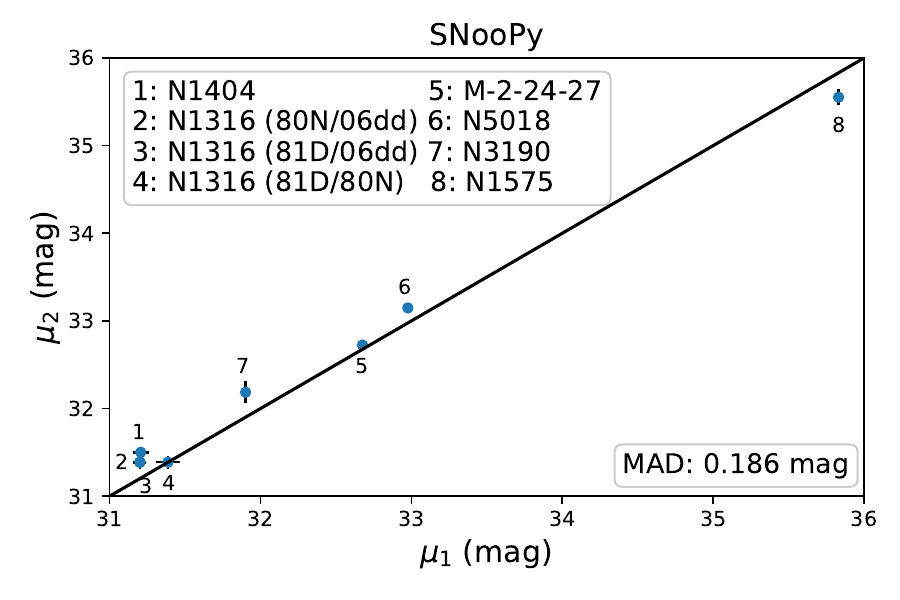}
    \label{fig:mu vs mu snoopy}\qquad
}
\caption{Comparing measured $\mu$ values obtained from both LC fitting models. $\mu_{1}$ represents one sibling in the pair and $\mu_{2}$ represents the other --- either sibling could be on either axis. The black diagonal line represents y = x. Each pair is labeled with a number and indicated in the legend. The median absolute difference (MAD) for each set is also given. The uncertainties shown do not include intrinsic scatter.} 
\label{fig: mu vs mu}
\end{figure}

We provide a list of all $\mu$ values from both SALT3-NIR and SNooPy in Table~\ref{tab:muvalues}. 
The uncertainties in distance moduli do not include expected uncertainty from intrinsic scatter, as this is a parameter we aim to measure.
With six sets of siblings, one of which is a set of four SNe and five of which are sibling pairs, this provides 11 potential $\mu$ comparisons. 
Although \citet{Hoogendam2022} demonstrate that two 91bg-like siblings provide consistent $\mu$ estimates, we find, like \citet{Stritzinger2010}, that the derived distance modulus from SN~2006mr is much larger than the other three siblings in NGC 1316.
\citet{Burns2018b} find consistency between the distance moduli of SN~2006mr and its siblings, reporting an $s_{BV}$ value of 0.26. However, we are unable to recover a $\mu$ value for SN~2006mr that is within 0.5 mag of its siblings even when we fix $s_{BV}$ to this value.
Thus, we do not include SN~2006mr in our sample of $\mu$ comparisons.
This reduces the total number of $\mu$ comparisons to eight. 
We plot this final sample comparing the $\mu$ values between siblings in Fig.~\ref{fig: mu vs mu}.
$\mu$ uncertainties do not include expected uncertainty from intrinsic scatter, and SALT3-NIR $\mu$ uncertainties are smaller than those from SNooPy in large part because they do not include uncertainties due to stretch and color corrections.

Using the SALT3-NIR model, we find a median absolute difference for the eight siblings $\mu$ comparisons of 0.248~mag and a range of absolute differences from 0.007 to 0.747~mag. Using the SNooPy model, we find a median absolute difference of 0.186~mag and a range from 0.001 to 0.295~mag.
As can be seen from Fig.~\ref{fig:salt3 lightcurves}, the SN~2002cv and SN~2002bo pair have poor LC fits from SALT3-NIR which are slightly improved when using SNooPy. This pair also corresponds to the largest $\mu$ residual of 0.747~mag from SALT3-NIR. 
The SNooPy fits result in more consistent $\mu$ values between siblings than the SALT3-NIR fits, however, we recognize that the SNooPy fits incorporate fits to stretch and color parameters while the SALT3-NIR fits do not.

When calculating SALT3-NIR fits, we fix both $x_1$ and $c$ to zero, treating the SNe as standard candles.
When attempting to fit for stretch and color parameters with SALT3-NIR and NIR data alone, the fits are unable to be constrained using \texttt{SNANA}.
Furthermore, we cannot use optically-fitted $x_1$ and $c$ parameters because optical data are not available for all SNe in our sample.
When we try fixing $A_V=0$~mag and $s_{BV}=1$ with SNooPy, rather than fitting for $A_V$ and $s_{BV}$, in a similar fashion to fixing both $x_1$ and $c$ for SALT3-NIR, we find that the SNooPy fits become worse and the differences in $\mu$ values between siblings increase; the median absolute difference in $\mu$ values goes from 0.186~mag to 0.279~mag.

Distance moduli using NIR data for some pairs of SN siblings have been analyzed previously in the literature. \citet{Stritzinger2010} report the range of differences in distance modulus is 0.236~mag for SN~1980N, SN~1981D, and SN~2006dd using SNooPy and NIR data alone, with SN~2006dd having the smallest $\mu$ value, and \citet{Gall2018} report a difference in $\mu$ values of 0.40~mag for SN~2007on and SN~2011iv (with SN~2007on having the larger $\mu$ value) using SNooPy and optical+NIR data and 0.20~mag with $H$-band alone. Results from our analysis are in agreement with these findings, particularly when using SNooPy. For SN~1981D, SN~1980N, and SN~2006dd, our $\mu$ values span 0.188~mag when using SNooPy and SN~2006dd results in the smallest $\mu$ value as well.
For SN~2007on and SN~2011iv, we find a difference in $\mu$ values of 0.30~mag and SN~2007on demonstrates the larger $\mu$ value.

\begin{table}[!hbt]
\centering
\tablenum{2}
\caption{Resulting distances from SALT3-NIR and SNooPy for all siblings using NIR data.} 

\begin{tabular}{cccc}
\hline \hline
SN Name & Host Galaxy & $\mu_\textrm{SALT3-NIR}$ & $\mu_\textrm{SNooPy}$ \\
\hline
\hline
1980N & \multirow{4}{*}{$\rm{NGC}~1316$} & 31.753(27) & 31.387(80) \\
1981D & & 31.759(28) & 31.389(73) \\
2006dd & & 31.466(22) & 31.201(44) \\
2006mr & & 32.889(19) & 33.039(100) \\
\hline
2002bo & \multirow{2}{*}{$\rm{NGC}~3190$} & 32.408(47) & 32.186(124) \\
2002cv & & 33.155(24) & 31.902(22) \\
\hline
2002dj & \multirow{2}{*}{$\rm{NGC}~5018$} & 33.187(19) & 33.148(44) \\
2021fxy & & 32.978(26) & 32.977(31) \\
\hline
2007on & \multirow{2}{*}{$\rm{NGC}~1404$} & 32.043(10) & 31.502(32) \\
2011iv & & 31.343(17) & 31.207(53) \\
\hline
2011at & \multirow{2}{*}{MCG -2-24-27} & 32.813(24) & 32.724(62) \\
2020jgl & & 32.669(30) & 32.676(32) \\
\hline
2020sjo & \multirow{2}{*}{$\rm{NGC}~1575$} & 35.780(22) & 35.830(38) \\
2020zhh & & 35.597(56) & 35.552(94) \\
\hline
\label{tab:muvalues}
\end{tabular}
\end{table}

\section{Comparison to Simulations}\label{sec: compare sim}
To estimate agreement between SN siblings, we find LCs from a simulated sample of SNe Ia which have similar redshift values, within 0.005 in redshift, and compare the results to our sample's results following \citet{Scolnic2020}. 
We construct these simulations of NIR LCs following Peterson et al.~in prep.~who use \texttt{SNANA} to generate accurate simulations based on characteristics from DEHVILS survey data such as LC cadence, coverage, and quality. 
While our sample of SNe is only $\sim30\%$ from DEHVILS, these simulations provide a baseline set of LCs from which we can choose LCs with similar redshifts as the LCs from our sample.
To evaluate the amount of intrinsic scatter present in our sample, simulations are created both with and without intrinsic scatter included (0.14 mag). This scatter is modeled as achromatic, similar to that done for the optical in \citet{Guy2010}. All simulated LCs are fit using SALT3-NIR with both $x_1$ and $c$ fixed to zero.
Given that minimal work has been done on creating simulations following the SNooPy model, we use the SALT3-NIR model here in this work for the construction of simulations, however we encourage future works to incorporate SNooPy simulations.

\begin{figure}[!tb]
    \centering
    \includegraphics[width=\columnwidth]{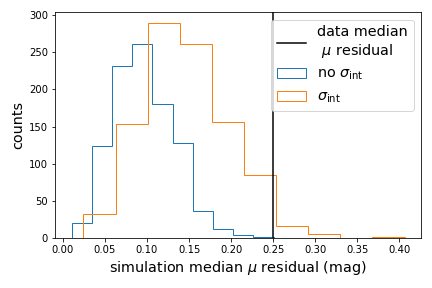}
    \caption{Each histogram features statistics from 8,000 random simulation differences in $\mu$ residuals obtained from lists of SNe that have redshift values within $0.005$ of each of our SN siblings. The blue histogram includes simulated data with $\sigma_{\rm int}$ while the orange histogram does not include $\sigma_{\rm int}$. One count on this histogram is the median of the list of eight simulated differences in $\mu$ residuals. 99.9\% of simulated sample without $\sigma_{\rm int}$ has differences lower than our differences while 97.3\% of the simulated sample with $\sigma_{\rm int}$ has differences lower than our differences. Our differences are represented by the vertical dotted line.}
    \label{fig: simulation}
\end{figure}

We construct our first simulation, including intrinsic scatter, totaling 26,533 LCs. From this, we compile 1,000 sets of eight differences in $\mu$ values where each of the $\mu$ values (corrected to be at the same redshift) are selected at random from the sample of simulated SNe with redshift values within 0.005 of each of our SN siblings' redshift values. We present a histogram of the median absolute differences in $\mu$ values for those 1,000 sets of eight differences in $\mu$ values in Fig.~\ref{fig: simulation} in blue, and indicate the median absolute difference from SALT3-NIR for our sample as a dotted black vertical line. If there were agreement between our sample's median absolute difference and the simulation median absolute differences in $\mu$ values, the line in Fig.~\ref{fig: simulation} would lie closer to the peak of the blue histogram. In the figure, we can see that our data value lies on the right side. 2.7\% of this sample has differences greater than the differences in our data sample. The standard deviation of the median absolute differences in $\mu$ values from this simulation, with intrinsic scatter, of 0.052 mag provides an estimate for uncertainty on the median absolute differences in $\mu$ values in our analysis.

We then make another simulation in which we do not include intrinsic scatter, which can be seen as the dotted orange histogram in Fig.~\ref{fig: simulation}.
The peak of the orange histogram falls to the left of the sample with intrinsic scatter, as expected, and we find the resulting median absolute difference in $\mu$ values from our sample to be even more unlikely if intrinsic scatter were not present in SNe Ia.
Approximately 0.1\% of the simulated values, not accounting for intrinsic scatter, have differences greater than the differences in our data sample. \textit{We can now show from this SN siblings analysis, with high significance, that there must be intrinsic scatter present in the NIR that cannot be attributed to host galaxy properties.}

\section{Discussion and Conclusion}\label{sec: discuss and conclude}
We perform a complete literature search of all SNe discovered and find six sets of SN siblings with NIR data.
Of these six sibling sets, one set includes four SNe (three of which are Type Ia-normal), and one SN LC fails quality cut criteria.
This results in a total of eight pairs of SN siblings with median absolute differences in $\mu$ values of 0.248~mag and 0.186~mag when fitting with SALT3-NIR and SNooPy, respectively.
Utilizing simulations, we find SN siblings seem to be no closer in terms of $\mu$ than two random SNe at the same redshift.
When the simulations do not include intrinsic scatter, recovering differences of the same scale becomes even more unlikely, occurring only $0.1\%$ of the time.
While we do not attribute all of the scatter observed in the data to $\sigma_{\rm int}$, this finding supports the existence of intrinsic scatter in the NIR.

With predominantly optical data alone, \citet{Burns2020} report a level of consistency in distance of $\sim3 \%$; for our NIR sample, we report approximately 9\% consistency in distance. To better understand this discrepancy, we obtain optical-only LCs from ATLAS and Pantheon+ for six out of eight pairs of siblings used in this analysis which, when fit with SALT3 and fitting for $x_1$ and $c$, have a median absolute difference in $\mu$ values of 0.177~mag.
The median absolute differences from the same set of six pairs with NIR-only LCs are 0.248~mag and 0.186~mag from using SALT3-NIR and SNooPy, respectively.
While the optical data present smaller differences in $\mu$ values between siblings, it is possible that this difference is due to the low statistics, data quality, and data reduction of this NIR sample.

One interesting finding about the SALT3-NIR model \citep{Pierel2022} is that even though the photometric data demonstrate that there are SNe that decline quickly even in the NIR, modifying the ``stretch" parameter $x_1$ (which is defined in the optical) does not result in reasonable fits for fast-declining NIR LCs.
This inability to fit fast-declining NIR LCs is most evident for SN~2007on where the characteristic secondary maximum in the NIR begins $\lesssim 20$ days past NIR maximum as compared to the more typical $\sim 30$ days past NIR maximum \citep{Dhawan2015,Mandel2022}.
Further, as can be seen from figure~7 in \citet{Pierel2022}, modifying SALT3 $x_1$ does not change the phase of any of the LC features for $J$- or $H$-band significantly at all.
In contrast, SNooPy's $s_{BV}$ parameter does act as a ``stretch" parameter in the NIR and by fixing $s_{BV}=0.57$ as derived by \citet{Gall2018}, we obtain a reasonable fit to the LC of SN~2007on from SNooPy. 

In this analysis, we fix both $x_1$ and $c$ when using SALT3-NIR. When we attempt to fit for $x_1$ and $c$ using SALT3-NIR and NIR data alone, \texttt{SNANA} is unable to produce LC fits for half of the SNe. All other SN LCs successfully fit by \texttt{SNANA} result in either unreasonable $x_1$ or $c$ parameters ($|x_1|>3$, $|c|>0.3$) or unreasonable uncertainties on those parameters ($\sigma_{x_1}>1$, $\sigma_c>0.1$).

We encourage further analysis into NIR LC models not just for SALT3-NIR, but also for SNooPy. Particularly, we believe improvements could be made in fitting the secondary maximum since we find the secondary maxima in our sample to be fit relatively poorly (e.g., SN~2011iv, SN~2002dj, SN~2002cv). We expect improvements could be made by adding to the training samples for both SALT3-NIR and SNooPy and specifically adding SNe with quality coverage of both the primary and secondary maxima across a large phase range. Interestingly, in the case of SN~2011at, which only has NIR data covering the secondary maximum, the LC fits well and the $\mu$ value is consistent with that from SN~2020jgl (only a 0.05~mag difference) indicating potential for measuring distances with NIR LCs even if the primary maximum is missed.

With the Roman space telescope \citep{Spergel2015, Hounsell2018,Rose2021}, $\sim$ 1,300 SNe will be observed in the first two years with rest frame NIR data (Peterson et al.~in prep.).
Given that \citet{Scolnic2020} report eight sibling pairs out of $\sim$ 3,000 SNe, we can expect approximately 3--4 more NIR sibling pairs from Roman.
This is not a sizeable increase in statistics, but it will be an interesting cross-check for Roman nonetheless. 
Additionally, if we search through NIR samples with larger statistics but less LC coverage, such as the Hawaii Supernova Flows survey (HSF; Do et al.~in prep.), we may find more siblings that could be used in a future NIR analysis.
For now, the next step is to understand whether or not the advantages to SN observations in the NIR can be seen with SN siblings from a larger sample NIR LCs.

\begin{acknowledgements}
This research has made use of the NASA/IPAC Extragalactic Database (NED), which is funded by the National Aeronautics and Space Administration and operated by the California Institute of Technology. UKIRT is owned by the University of Hawaii (UH) and operated by the UH Institute for Astronomy. When (some of) the data reported here were obtained, the operations were enabled through the cooperation of the East Asian Observatory.
D.S. is supported by Department of Energy grant DE-SC0010007, the David and Lucile Packard Foundation, the Templeton Foundation and Sloan Foundation. 
This research has made use of NASA’s Astrophysics Data System.

Software:
\texttt{SNANA} \citep{Kessler2009}, astropy \citep{AstropyCollab2013, Price-Whelan2018}, matplotlib \citep{Hunter2007}, NumPy \citep{vanderWalt2011}
\end{acknowledgements}

\bibliographystyle{mn2e}
\bibliography{main}{}

\appendix
\section{Photometric Calibration Analysis}\label{appendix_photometric_analysis}

Here, we ensure the calibration of the data is relatively consistent and accurate enough for use in our analysis. We do this by comparing the individual band magnitudes from stellar photometry to 2MASS magnitudes. In Fig.~\ref{fig: appendix 3}, each subfigure features a residual mag vs.~mag plot. These plots show the difference between the 2MASS stellar magnitudes and the respective magnitudes noted in the respective paper as a function of magnitude. The black dotted lines are the median of those residuals, which is given in the legend of each subfigure. To remove outliers, we placed a cut requiring the absolute value of the residuals to be $<$ 0.5~mag. The median of all the residuals combined is 0.008~mag. Additionally, the range of median magnitude residuals is from $-$0.045 to 0.130~mag.
When we do not consider the largest of the median magnitude residuals of 0.130~mag and 0.080~mag from comparing stellar photometry taken over 40 years ago to 2MASS, the range of median magnitude residuals goes from $-$0.045 to 0.025~mag.
If we take this offset into account by adding $\sim$0.1~mag to the $\mu$ values obtained for SN~1980N and SN~1981D, this actually increases the difference between the $\mu$ values of these SNe and that of SN~2006dd.

\newcommand{\appenwidth}{0.47}
\begin{figure*}[!ht]
\subfloat{
    \includegraphics[width=\appenwidth\textwidth]{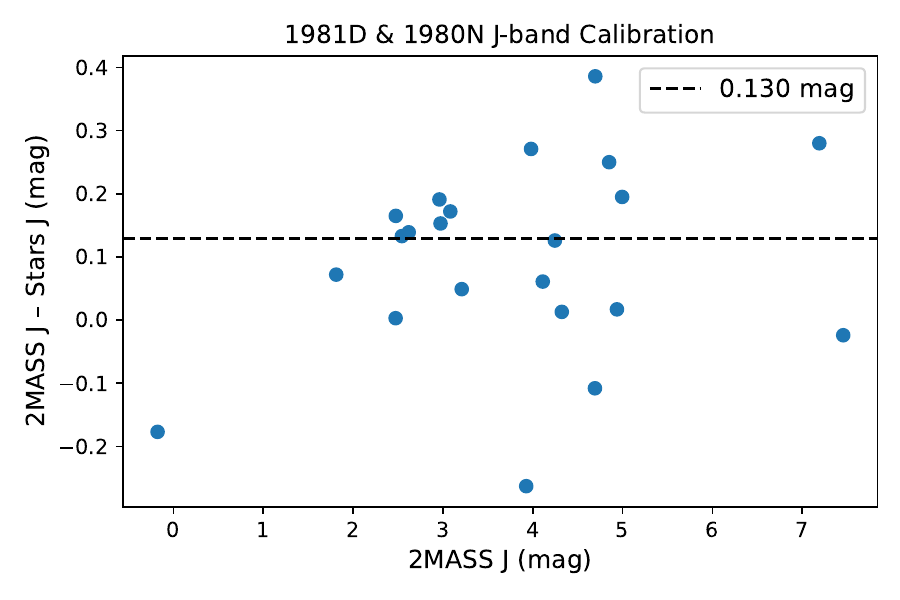}
    \label{fig:1}\qquad
}
\hspace{-1cm}
\subfloat{
    \includegraphics[width=\appenwidth\textwidth]{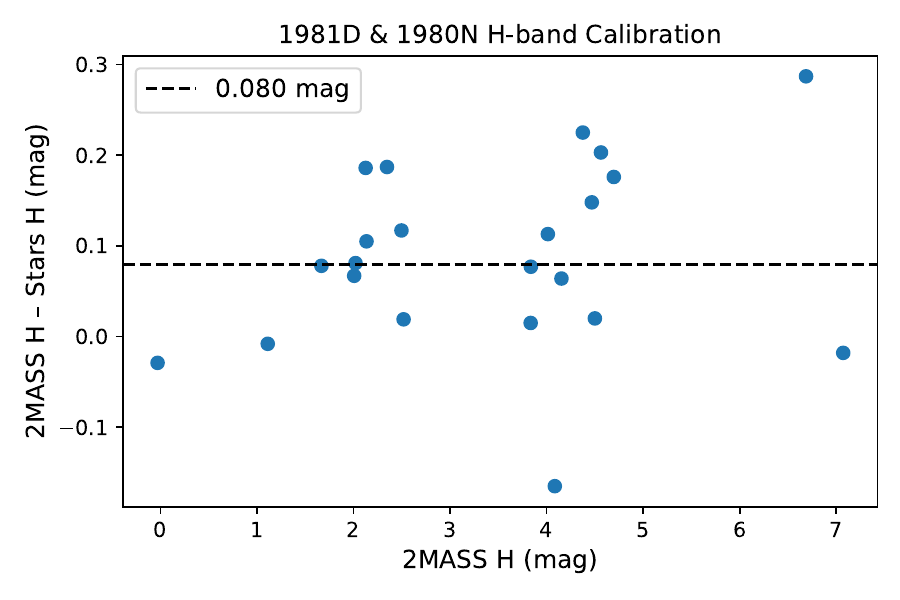}
    \label{fig:2}\qquad
}
\\
\subfloat{
    \includegraphics[width=\appenwidth\textwidth]{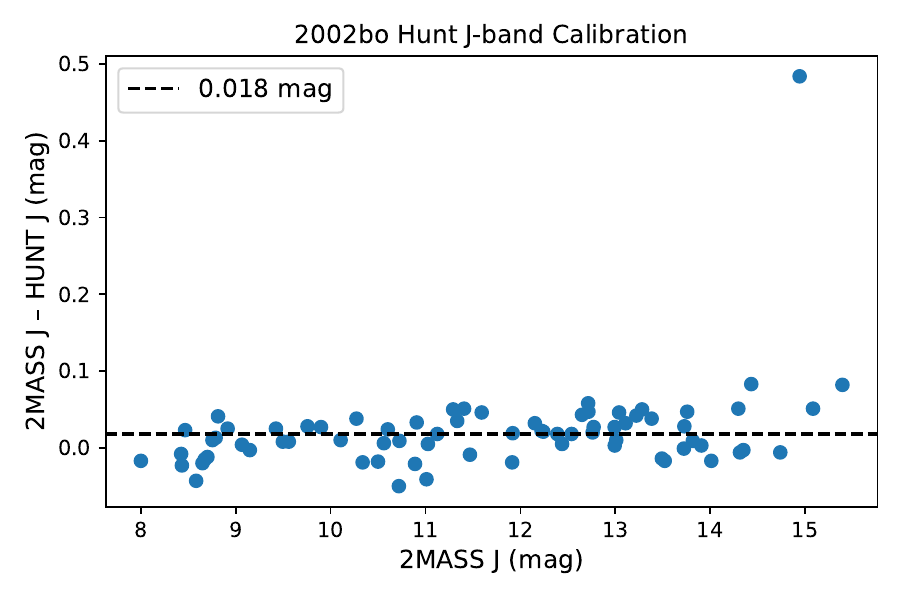}
    \label{fig:3}\qquad
}
\hspace{-0.9cm}
\subfloat{
    \includegraphics[width=\appenwidth\textwidth]{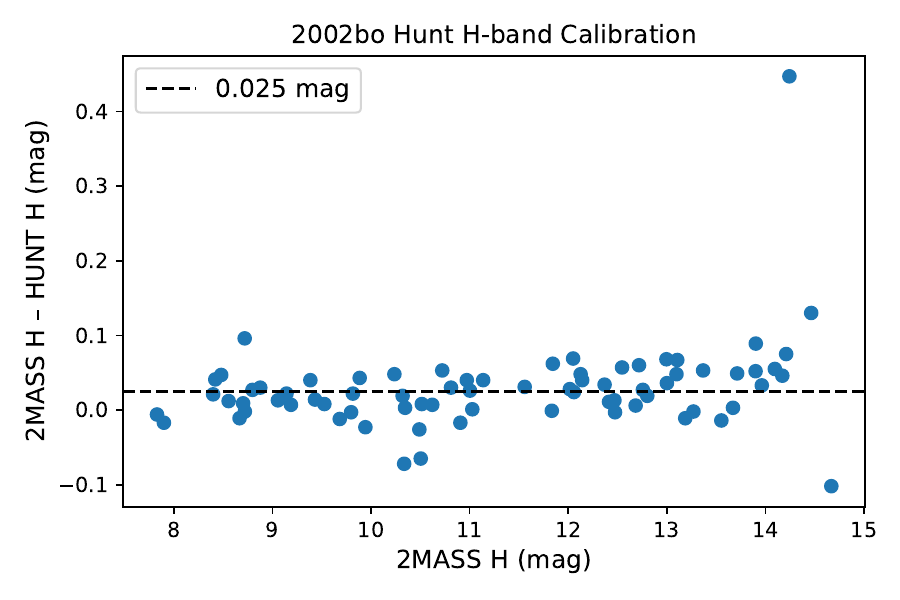}
    \label{fig:4}\qquad
}
\label{fig: appendix}
\end{figure*}

\begin{figure*}
\subfloat{
    \includegraphics[width=\appenwidth\textwidth]{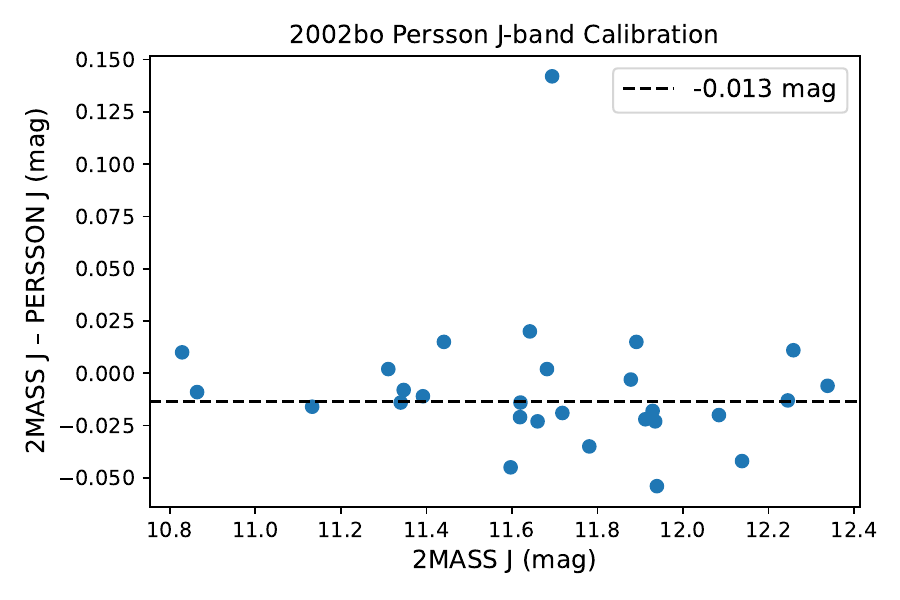}
    \label{fig:5}\qquad
}
\hspace{-0.9cm}
\subfloat{
    \includegraphics[width=\appenwidth\textwidth]{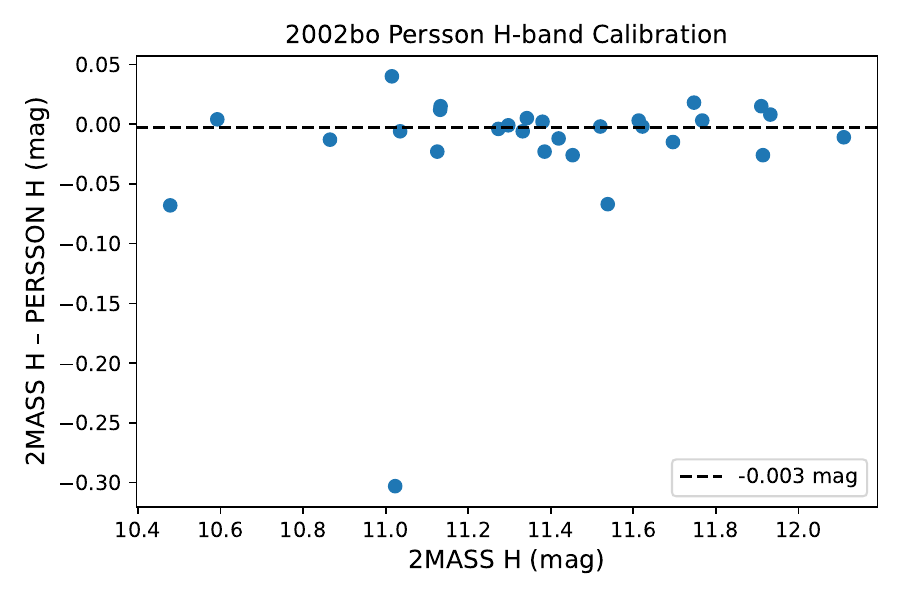}
    \label{fig:6}\qquad
}
\\
\subfloat{
    \includegraphics[width=\appenwidth\textwidth]{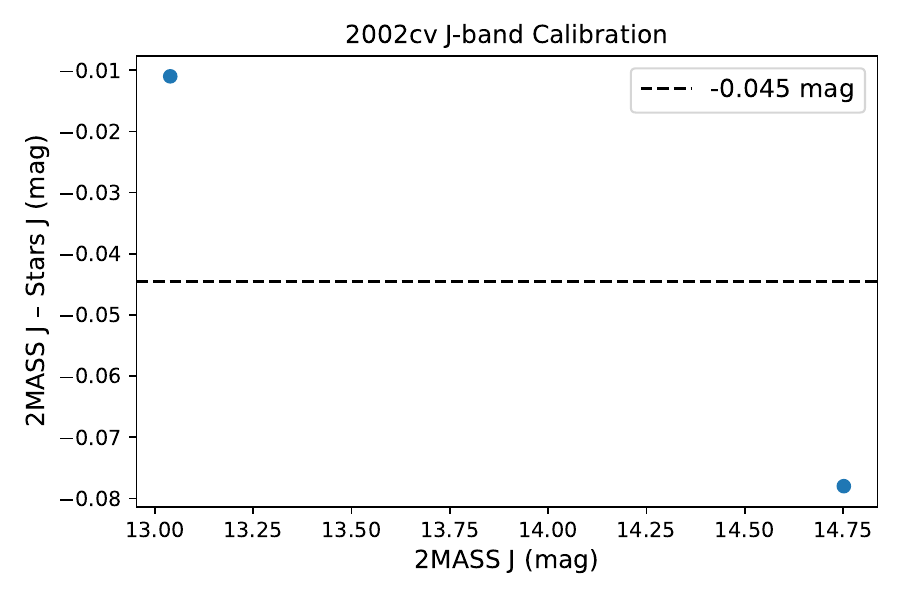}
    \label{fig:7}\qquad
}
\hspace{-0.9cm}
\subfloat{
    \includegraphics[width=\appenwidth\textwidth]{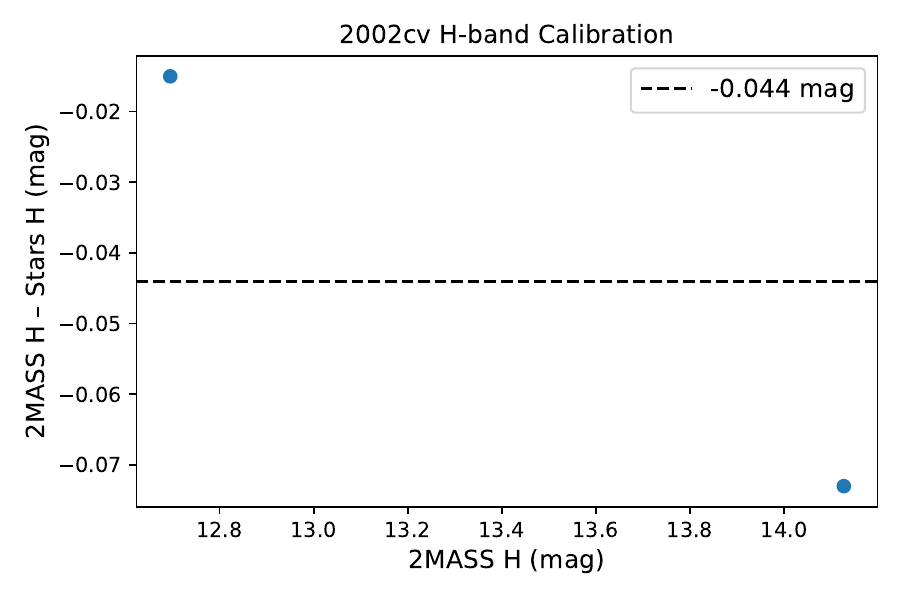}
    \label{fig:8}\qquad
}
\\
\subfloat{
    \includegraphics[width=\appenwidth\textwidth]{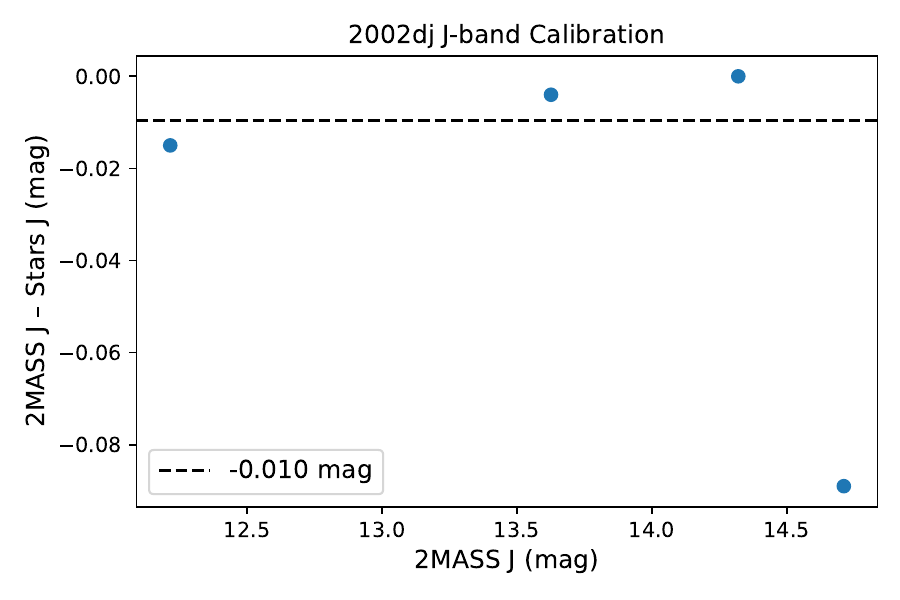}
    \label{fig:9}\qquad
}
\hspace{-0.9cm}
\subfloat{
    \includegraphics[width=\appenwidth\textwidth]{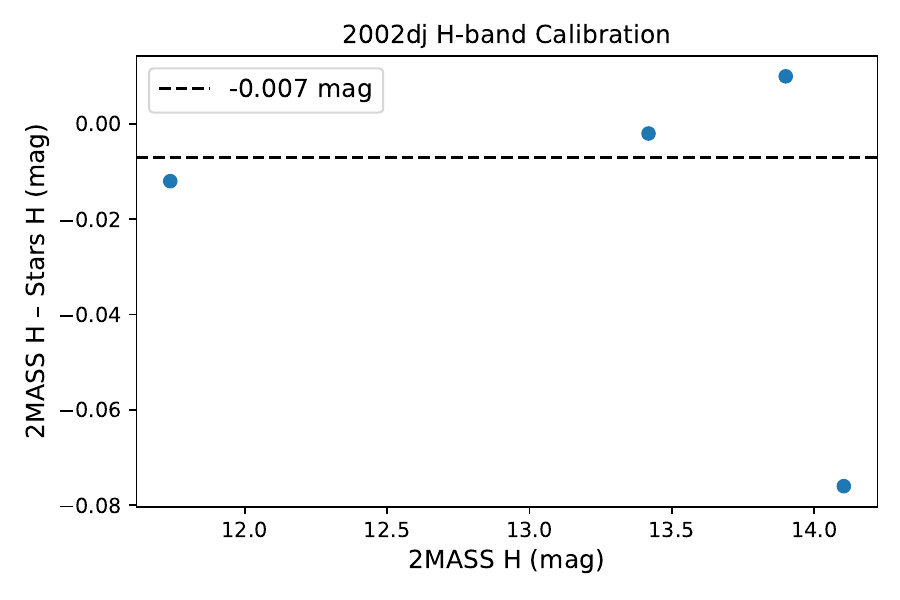}
    \label{fig:10}\qquad
}
\\
\subfloat{
    \includegraphics[width=\appenwidth\textwidth]{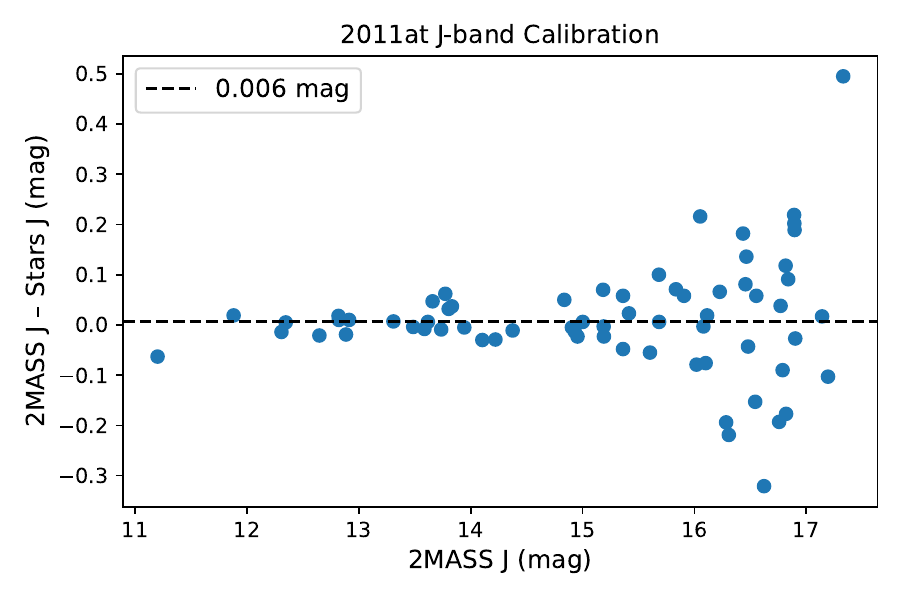}
    \label{fig:11}\qquad
}
\hspace{-0.8cm}
\subfloat{
    \includegraphics[width=\appenwidth\textwidth]{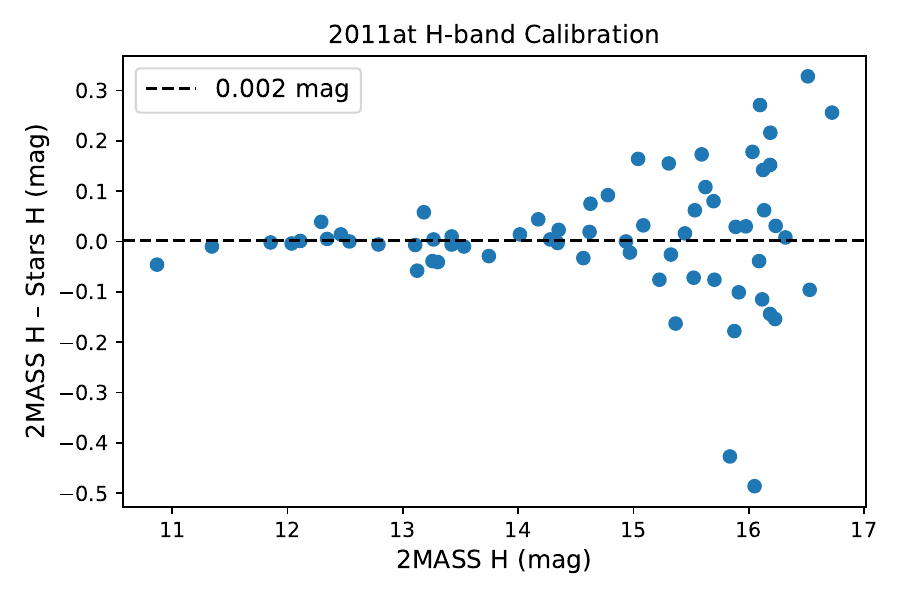}
    \label{fig:12}\qquad
}
\caption{Magnitude residuals versus magnitude plots comparing sets of stellar photometry for each SN sibling to 2MASS photometry.} 
\label{fig: appendix 3}
\end{figure*}

\end{document}